\documentclass[aps,pra,english,twocolumn,showpacs,preprintnumbers,amsmath,amssymb,floatfix]{revtex4-1}
\usepackage[T1]{fontenc}
\usepackage[latin9]{inputenc}
\usepackage{graphicx}
\usepackage{amssymb}

\usepackage{babel}
\makeatletter

\usepackage{esint}

\makeatother

\usepackage{babel}
\makeatother

\begin{document}

\title{Magnetoresistance of doped silicon}

\author{Antonio Ferreira da Silva }

\affiliation{Instituto de Fisica, Universidade Federal da Bahia, Campus Ondina 40210 340 Salvador, BA, Brazil}
 
\author{Alexandre Levine }

\affiliation{Instituto de Fisica, Universidade de S\~{a}o Paulo
Laborat\'{o}rio de Novos Materiais Semicondutores
05508-090 Butant\~{a}, S\~{a}o Paulo, SP, Brazil}

\author{Zahra Sadre Momtaz}

\affiliation{Instituto de Fisica, Universidade de S\~{a}o Paulo
Laborat\'{o}rio de Novos Materiais Semicondutores
05508-090 Butant\~{a}, S\~{a}o Paulo, SP, Brazil}

\author{Henri Boudinov}

\affiliation{Instituto de Fisica, Universidade Federal do Rio Grande do Sul
91501 970  Porto Alegre, Rio Grande do Sul , RS, Brazil}

\author{Bo E. Sernelius}
\email{bos@ifm.liu.se}
\affiliation{Division of Theory and Modeling, Department of Physics, Chemistry
and Biology, Link\"{o}ping University, SE-581 83 Link\"{o}ping, Sweden}

\begin{abstract}
We have performed longitudinal magnetoresistance measurements on heavily $n$-doped silicon for donor concentrations exceeding the critical value for the metal-non-metal transition. The results are compared to those from a many-body theory where the donor-electrons are assumed to reside at the bottom of the many-valley conduction band of the host. Good qualitative agreement between theory and experiment is obtained. 
\end{abstract}


\maketitle

\section{\label{Int}Introduction}
Magnetoresistance, the property of a material that its electrical resistivity changes when exposed to an external magnetic field, was first discovered by Lord Kelvin\,\cite{Kel}. More recent discoveries are the giant magnetoresistance (GMR)\,\cite{Fert,Grun}, the colossal magnetoresistance (CMR)\,\cite{Jonk,Ram}, the tunnel magnetoresistance (TMR)\,\cite{Jull}, the extraordinary magnetoresistance (EMR)\,\cite{Sol}, and  the large magnetoresistance (LMR)\,\cite{Ali}. All these spectacular properties were found in either magnetic systems or in special geometrical structures. 

Even in ordinary non-magnetic bulk materials there are interesting magnetoresistance effects. For all conducting pure single crystals it is experimentally found that the application of a magnetic induction $\bf B $ results in an increase of the resistivity $\rho $; the magnetoresistance ratio, or just magnetoresistance, defined as $\Delta \rho /\rho  = \left[ {\rho \left( B \right) - \rho \left( 0 \right)} \right]/\rho \left( 0 \right)$ is positive. This general behavior of the crystalline state is in sharp contrast to the conduction properties of a number of heavily doped semiconductors where one observes a negative magnetoresistance. There are many different models\,\cite{Yam,Sas,Ion,And,Zav,Emel,Ish,Kho} in the literature trying to explain this anomalous behavior. They are all related to a model by Toyozawa\,\cite{Toy} where the conduction electrons scatter against localized spins. We refer the reader to the review article by Alexander and Holcomb\,\cite{Alex} for the discussion of some of these models. The discussion is organized  around a model which includes three main features: above the critical donor concentration, $n_c$, the electrons are delocalized;  above a second critical donor concentration, $n_{cb}$, the Fermi level  passes into the conduction band of the host crystal; for $n_c<n_d<n_{cb}$ the electrons exist in a poorly understood ``impurity band'' leading to anomalous propeties. We proposed a different description\,\cite{SerBer} where the donor electrons end up in the conduction band of the host already at the critical concentration $n_c$. The anomalous properties on the metallic side of and close to the transition point we suggested were caused by many-body effects. Examples of anomalous behavior is that the resistivity, the heat capacity, and the spin susceptibility are all enhanced close to $n_c$. Another example is the negative magnetoresistance that we treat in this work.

Lately much research has been devoted to systems with positive magnetoresistance showing a linear dependence on the applied magnetic field\,\cite{John,Por}. 
Several mechanisms have been suggested to explain this behavior from
geometrical\,\cite{Bran}, classical\,\cite{Par1,Par2,Hu}, quantum\,\cite{Abr1,Abr2}, and effective medium\,\cite{Gut1,Gut2} perspectives. 

In this work we focus on the magnetoresistance of heavily doped semiconductors near and on the metallic side of the metal-non-metal transition.

The material is arranged in the following way. In Sec. \ref{Exp} we present the experimental details. Sec. \ref{The} is devoted to the theoretical model and derivations. Our experimental and theoretical results are compared in Sec. \ref{Res}. Finally, Sec. \ref{Sum} is a brief summary and conclusion section.

\section{\label{Exp}Experimental  Details}

\begin{figure}
\includegraphics[width=8.0cm]{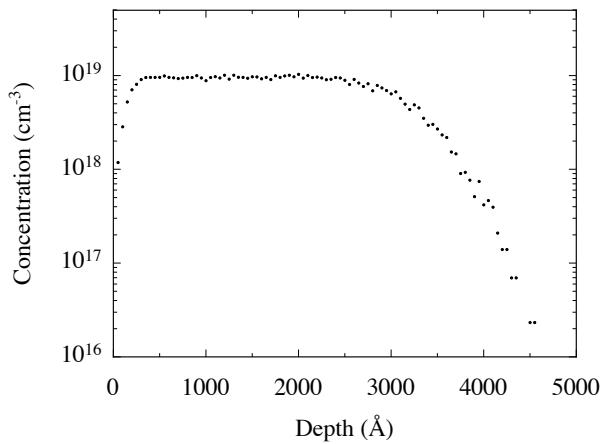}
\caption{Simulated concentration profile for nominal concentration $1 \times {10^{19}}{\rm{c}}{{\rm{m}}^{ - 3}}$.}
\label{figu1}
\end{figure}

We use p-type, (100)-oriented Si wafers with resistivity in the range of 16-25 $\Omega {\rm{cm}}$. Phosphorous was implanted at room temperature. Six implantations with energies of 180, 120, 80, 55, 30 and 15 keV were accumulated in each sample with proper doses to result in a plateau like profile of P from the surface to the depth of 0.30$\mu {\rm{m}}$ with $ \sim 5\% $ deviation, according to TRIM code simulation\,\cite{Zieg}. The implanted P doses were $1.4 \times {10^{14}} {\rm{c}}{{\rm{m}}^{ - 2}}$ (at 180 keV), $5.4 \times {10^{13}}{\rm{c}}{{\rm{m}}^{ - 2}}$ (at 120 keV), $3.6 \times {10^{13}}{\rm{c}}{{\rm{m}}^{ - 2}}$ (at 80 keV), $3.0 \times {10^{13}} {\rm{c}}{{\rm{m}}^{ - 2}}$ (at 55 keV), $2.2 \times {10^{13}}{\rm{c}}{{\rm{m}}^{ - 2}}$ (at 30 keV), and $1.2 \times {10^{13}}{\rm{c}}{{\rm{m}}^{ - 2}}$ (at 15 keV) in order to achieve a P concentration of $1 \times {10^{19}}{\rm{c}}{{\rm{m}}^{ - 3}}$. The simulated concentration profile is shown in Fig.\,\ref{figu1}. The doses in the other samples were scaled to this sample, according to the ratio of the desired P concentration. Samples with implanted P concentrations of $2 \times {10^{18}}{\rm{c}}{{\rm{m}}^{ - 3}}$, $4 \times {10^{18}}{\rm{c}}{{\rm{m}}^{ - 3}}$, $6 \times {10^{18}}{\rm{c}}{{\rm{m}}^{ - 3}}$, $8 \times {10^{18}}{\rm{c}}{{\rm{m}}^{ - 3}}$, and $1 \times {10^{19}}{\rm{c}}{{\rm{m}}^{ - 3}}$ were prepared. The damage annealing and the electrical activation of P were performed at 900 $^\circ {\rm{C}}$ for 20 minutes in argon atmosphere in a conventional furnace. 
Van der Pauw structures\,\cite{Pauw} were fabricated by manually applied indium contacts at the corners of square ${\rm{6}} \times {\rm{6 }}\, {{\rm{mm}}^{\rm{2}}}$ samples. Annealing at 300 $^\circ {\rm{C}}$ on a hot plate was performed to improve the contacts. The implantation process as well as the obtained values of the electron concentrations are also described in detail in Refs.\,\cite{Ferr} and\,\cite{Abram}.

	We performed magneto-transport measurements on the described structures with Van der Pauw geometries, exploiting conventional lock-in technique with frequencies 7 - 13 Hz, in the temperature range of  1.5-4.2 K  and bias current of 10 $\mu A$ which is low enough to prevent heating effect and at the same  time provide a well defined signal for our measurements. Both Hall and longitudinal resistance measurements were done in an Oxford cryostat with VTI (Variable Temperature Insert), in the presence of perpendicular magnetic field provided by a superconducting coil, capable of generating fields up to 12 T in ${}^4{\rm{He}}$ refrigerator. 
	

\section{\label{The}Theory}
\begin{figure}
\includegraphics[width=8.0cm]{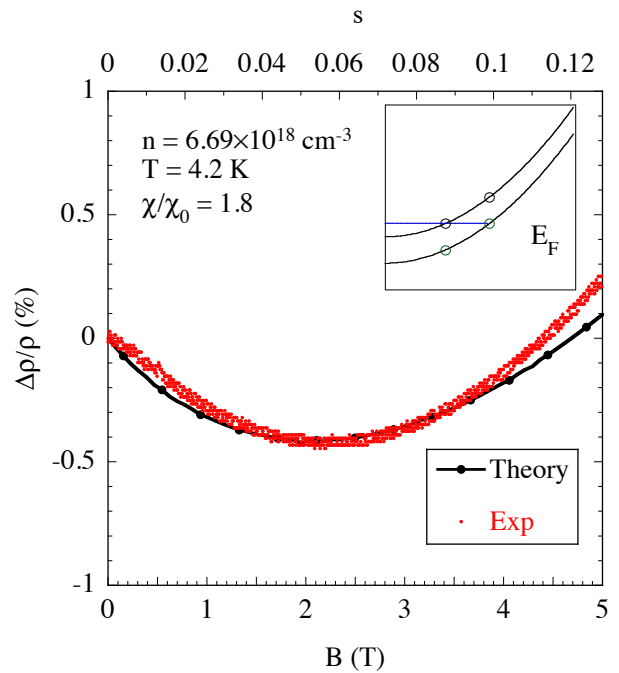}
\caption{(Color online)The magnetoresistance at 4.2 K as a function of magnetic induction, B, for a Si:P sample with doping concentration $6.69\times10^{18}$ cm$^{-3}$.  The red dots denote our experimental result and the solid curve with solid circles our theoretical result. The upper horizontal axis shows the spin-polarization parameter $s$. The inset shows schematically the energy dispersion of the spin up and spin down bands in presence of a magnetic field. The circles indicate which states are involved in the enhancement of the density of states. See the text for details.}
\label{figu2}
\end{figure}

We start with our approximations and notation. Si is a  semiconductor with $\nu  = 6$ anisotropic conduction band valleys. For heavily $n$-type doped silicon, on the metallic side of the metal-non-metal transition ($n > {n_c}$), the donor electrons are up in the conduction band valleys. The anisotropy has some effects on the resistivity\,\cite{Ser1} but we neglect this here and let the electrons be distributed in $\nu$ Fermi-spheres. The relation between the Fermi wave vector,  ${k_0}$, and the doping density, $n$, is given by
\begin{equation}
{k_0} = {\left( {3{\pi ^2}n/\nu } \right)^{1/3}}.
\label{equ1}
\end{equation}
The Fermi energy is
\begin{equation}
{E_0} = {\hbar ^2}k_0^2/\left( {2m} \right) = {\hbar ^2}k_0^2/\left( {2{m_{de}}{m_e}} \right)
\label{equ2}
\end{equation}
where ${{m_e}}$ is the electron mass and the density of states effective mass for a Fermi sphere is ${m_{de}} = {\left( {{m_l}m_t^2} \right)^{1/3}} = .322$. Apart from the kinetic energy there are contributions from the interactions between the electrons (the exchange and correlation energy, ${E_{xc}}$) and from the interactions with the ionized-donor potentials (the band-structure energy, ${E_b}$). These interaction energies  lead to a deformation of the parabolic band dispersion and a modification of the density of states. This modification is important for the effects discussed in this work so we discuss the density of states here.

\begin{figure}
\includegraphics[width=8.0cm]{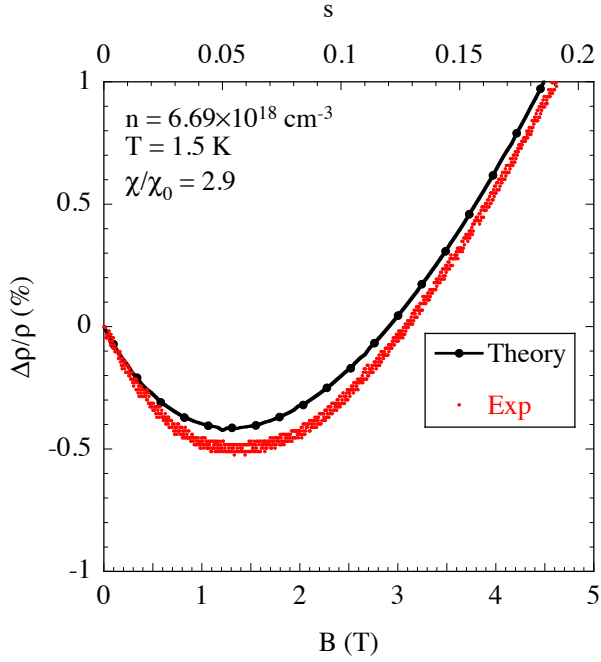}
\caption{(Color online) The same as Fig.\,\ref{figu2} but now for 1.5 K. }
\label{figu3}
\end{figure}

The density of states is the number of states per energy and volume. The density of states from one valley is
\begin{equation}
\begin{array}{l}
{D_E} = {D_k}/\left( {dE\left( k \right)/dk} \right) = \frac{{2 \cdot 4\pi {k^2}}}{{{{\left( {2\pi } \right)}^3}\left( {dE\left( k \right)/dk} \right)}}\\
\quad \quad  = \frac{{{k^2}}}{{{\pi ^2}\left( {dE\left( k \right)/dk} \right)}},
\end{array}
\label{equ3}
\end{equation}
where we have taken into account that in each valley there are two states for each ${\bf{k}}$, one with spin up and one with spin down. For non-interacting electrons the corresponding density of states is
\begin{equation}
D_E^0 = \frac{{{k^2}}}{{{\pi ^2}\left( {d{E^0}\left( k \right)/dk} \right)}} = \frac{{km}}{{{\pi ^2}{\hbar ^2}}}.
\label{equ4}
\end{equation}
We may express the density of states for interacting electrons on an analogous form by introducing a wave-number dependent effective mass,
\begin{equation}
{D_E} = \frac{{k{m^*}}}{{{\pi ^2}{\hbar ^2}}}.
\label{equ5}
\end{equation}
 The  effective mass can be written as
\begin{equation}
{m^*}\left( k \right) = m/\left[ {1 - \beta \left( k \right)} \right],
\label{equ6}
\end{equation}
where ${\beta}\left( k \right)$ gets a contribution from each of the interaction energies, ${\beta }\left( k \right) = \beta _{xc}\left( k \right) + \beta _b \left( k \right)$, where
\begin{equation}
\begin{array}{l}
\beta _{xc} \left( k \right) =  - \frac{m}{{{\pi ^2}k}}\frac{\partial }{{\partial k}}\frac{{\delta N \cdot {E_{xc}}}}{{\delta {n }\left( {\bf{k}} \right)}},\\
\beta _b \left( k \right) =  - \frac{m}{{{\pi ^2}k}}\frac{\partial }{{\partial k}}\frac{{\delta N \cdot {E_b}}}{{\delta {n }\left( {\bf{k}} \right)}}
\end{array}
\label{equ7}
\end{equation}
The quantity ${n }\left( {\bf{k}} \right)$ is the occupation number of the state with wave-vector ${\bf{k}}$, and $N$ is the total number of electrons. One effect of the interactions, that turns out to be very important for the present work, is that the effective mass and density of states are enhanced in a region around the Fermi level (see Fig.\,3 of Ref.\,\cite{Ser2}).

Before we continue let us introduce some dimensionless variables that we will use throughout,
\begin{equation}
\begin{array}{l}
Q = q/2{k_0},\\
W = \hbar \omega /4{E_0},\\
P = k/2{k_0},\\
y = \frac{{\nu m{e^2}}}{{{\hbar ^2}\kappa {k_0}}},
\end{array}
\label{equ8}
\end{equation}
where $\kappa =11.4$ is the background dielectric constant of Si.
In RPA (random phase approximation) the $\beta$-functions at the Fermi-level are
\begin{equation}
\begin{array}{l}
{\beta _{xc}}\\
 = \frac{y}{{\nu \pi }}\left\{ {\left[ {1 - \int\limits_0^1 {dQ\frac{1}{{Q\tilde \varepsilon \left( {Q,0} \right)}}} } \right. + \frac{1}{\pi }\int\limits_0^\infty  {dQ\int\limits_0^\infty  {dW\left[ {\frac{1}{{\tilde \varepsilon \left( {Q,iWQ} \right)}} - 1} \right]} } } \right.\\
\left. {\quad \quad  \times \left[ {\ln \left| {\frac{{{W^2} + {{\left( {Q + 1} \right)}^2}}}{{{W^2} + {{\left( {Q - 1} \right)}^2}}}} \right| - \frac{{2\left( {Q + 1} \right)}}{{{W^2} + {{\left( {Q + 1} \right)}^2}}} - \frac{{2\left( {Q - 1} \right)}}{{{W^2} + {{\left( {Q - 1} \right)}^2}}}} \right]} \right\},\\
{\beta _b} = \frac{{2{y^2}}}{{3\nu {\pi ^2}}}\int\limits_0^\infty  {dQ\frac{1}{{{Q^2}\left( {{Q^2} - 1} \right){{\tilde \varepsilon }^2}\left( {Q,0} \right)}}} \left[ {1 - \frac{{{Q^2} - 1}}{{2Q}}\ln \left| {\frac{{Q + 1}}{{Q - 1}}} \right|} \right],
\end{array}
\label{equ9}
\end{equation}
where $\tilde \varepsilon \left( {{\bf{q}},\omega } \right) = 1 + \alpha \left( {{\bf{q}},\omega } \right)$ is the dielectric function in RPA for an electron system of $\nu $  Fermi spheres embedded in Si with an effective dielectric constant, $\kappa $. In the expression for ${\beta _b}$ we have assumed that there are as many ionized donor potentials as there are electrons in the Fermi spheres and that these potentials can be represented by randomly distributed pure Coulomb potentials.

\begin{figure}
\includegraphics[width=8.0cm]{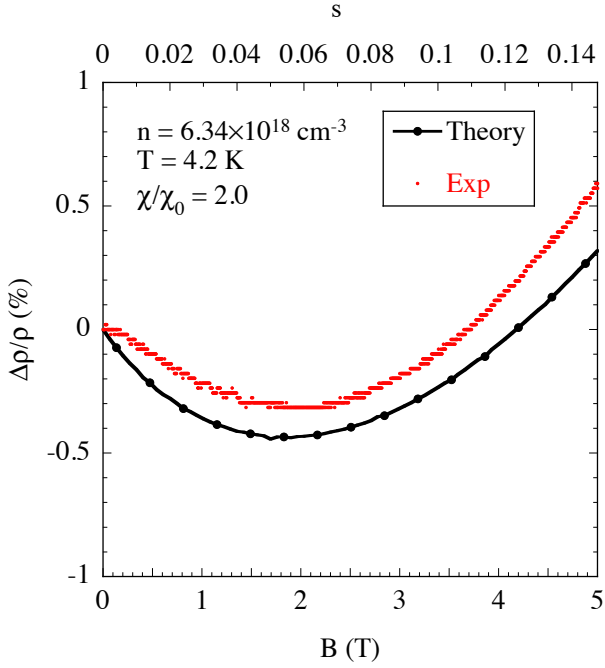}
\caption{(Color online) The same as Fig.\,\ref{figu2} but now for the doping concentration $6.34\times10^{18}$ cm$^{-3}$.}
\label{figu4}
\end{figure}

\begin{figure}
\includegraphics[width=8.0cm]{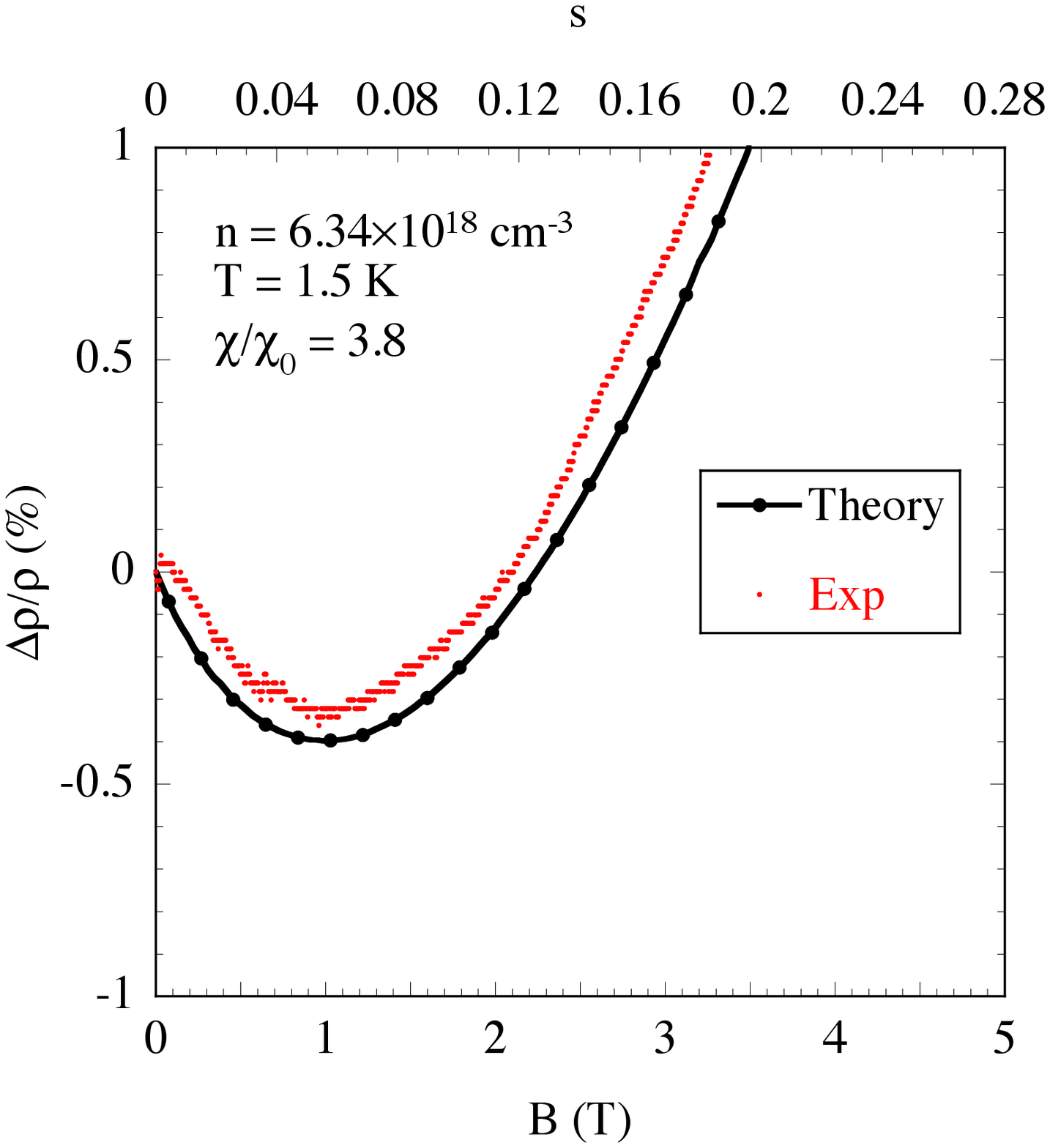}
\caption{(Color online) The same as Fig.\,\ref{figu3} but now for the doping concentration $6.34\times10^{18}$ cm$^{-3}$. }
\label{figu5}
\end{figure}
\begin{figure}
\includegraphics[width=8.0cm]{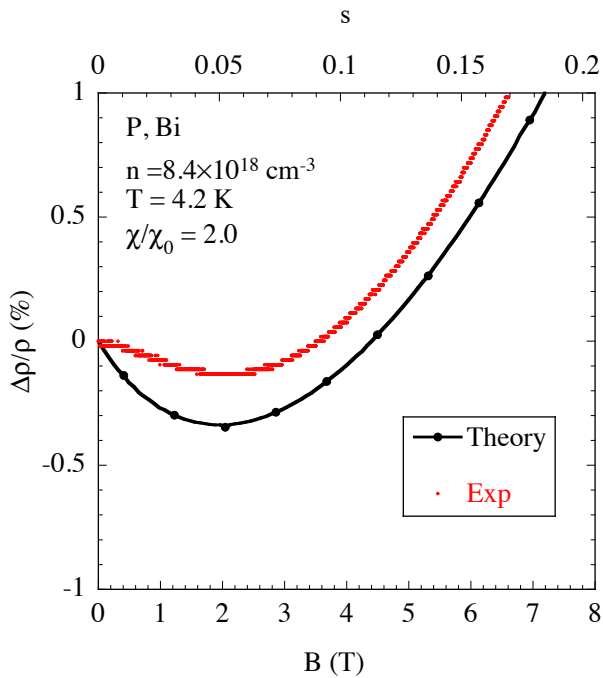}
\caption{(Color online) The same as Fig.\,\ref{figu3} but now for a sample with two donors, P and Bi, with the total doping concentration $8.4\times10^{18}$ cm$^{-3}$. }
\label{figu6}
\end{figure}

The analytical expressions for the polarizabilities needed in the calculations are
\begin{equation}
\begin{array}{l}
\alpha \left( {Q,iWQ} \right) = \frac{y}{{2\pi {Q^2}}}\left\{ {1 + \frac{{{W^2} + 1 - {Q^2}}}{{4Q}}\ln \left[ {\frac{{{W^2} + {{\left( {1 + Q} \right)}^2}}}{{{W^2} + {{\left( {1 - Q} \right)}^2}}}} \right]} \right.\\
\left. {\quad \quad \quad \quad \quad \quad  - W\left[ {{{\tan }^{ - 1}}\frac{{\left( {1 + Q} \right)}}{W} + {{\tan }^{ - 1}}\frac{{\left( {1 - Q} \right)}}{W}} \right]} \right\},
\end{array}
\label{equ10}
\end{equation}
and
\begin{equation}
\alpha \left( {Q,0} \right) = \frac{y}{{2\pi {Q^2}}}\left[ {1 + \frac{{1 - {Q^2}}}{Q}\ln \left| {\frac{{1 + Q}}{{1 - Q}}} \right|} \right].
\label{equ11}
\end{equation}

The resistivity we calculate by using the so-called generalized Drude approach\,\cite{SerMor,Ser3,Ser4}. In the static case which is what we need here the results agree with the so-called Ziman's formula\,\cite{Ziman}.
\begin{equation}
\begin{array}{l}
\rho  = \frac{1}{\sigma } = \frac{1}{{n{e^2}\tau /{m^*}}},\\
\frac{1}{\tau } = \frac{4}{3}\frac{{\nu {e^4}m}}{{\pi {\hbar ^3}{\kappa ^2}}}\int\limits_0^1 {dQ\frac{1}{{Q{{\tilde \varepsilon }^2}\left( {Q,0} \right)}}}, 
\end{array}
\label{equ12}
\end{equation}
where $\rho $, $\sigma $, and $\tau $ are the resistivity, conductivity and transport time, respectively.

When a static and spatially homogeneous magnetic field (magnetic induction {\bf B}) is applied the bands with spin up electrons (spin parallel to  {\bf B}) move up in energy and those with spin down electrons (spin antiparallel to  {\bf B}) move down. There is a redistribution of the electrons so that more electrons have spin down than have spin up. This has the effect that the density of states, the effective mass at the Fermi level, the contribution to the conductivity, and the transport time are no longer the same for the two groups of electron. Let us introduce the spin-polarization parameter, $s$, that varies from zero in absence of {\bf B} to $1$ at full polarization (all electrons have spin down),
\begin{equation}
s = \frac{{n \downarrow  - n \uparrow }}{n}.
\label{equ13}
\end{equation}

The density and Fermi wave-number of spin up and down electrons are

\begin{equation}
\begin{array}{l}
n \uparrow  = \frac{{1 - s}}{2}n,\\
n \downarrow  = \frac{{1 + s}}{2}n,\\
{k_0} \uparrow  = {k_0}/a,\\
{k_0} \downarrow  = {k_0}/b,
\end{array}
\label{equ14}
\end{equation}
where
\begin{equation}
\begin{array}{l}
a = {\left( {1 - s} \right)^{ - 1/3}},\\
b = {\left( {1 + s} \right)^{ - 1/3}}.
\end{array}
\label{equ15}
\end{equation}

The resistivity is now
\begin{equation}
\begin{array}{l}
\rho  = \frac{1}{{n \uparrow {e^2}\tau  \uparrow /{m^*} \uparrow  + n \downarrow {e^2}\tau  \downarrow /{m^*} \downarrow }}\\
\quad  = \frac{1}{{\frac{{n \uparrow {e^2}\tau  \uparrow \left( {1 - \beta  \uparrow } \right)}}{m} + \frac{{n \downarrow {e^2}\tau  \downarrow \left( {1 - \beta  \downarrow } \right)}}{m}}}\\
\quad  = \frac{{m/{e^2}}}{{n \uparrow \tau  \uparrow \left( {1 - \beta  \uparrow } \right) + n \downarrow \tau  \downarrow \left( {1 - \beta  \downarrow } \right)}}.
\end{array}
\label{equ16}
\end{equation}

The polarizability can be divided into a contribution from each group of electrons
\begin{equation}
\begin{array}{l}
\alpha \left( {Q,iWQ} \right) = \alpha  \uparrow \left( {Q,iWQ} \right) + \alpha  \downarrow \left( {Q,iWQ} \right),\\
\alpha  \uparrow \left( {Q,iWQ} \right) = \frac{a}{2}\alpha \left( {aQ,i{a^2}WQ} \right),\\
\alpha  \downarrow \left( {Q,iWQ} \right) = \frac{b}{2}\alpha \left( {bQ,i{b^2}WQ} \right),
\end{array}
\label{rqu17}
\end{equation}
and the new transport times become

\begin{equation}
\begin{array}{l}
\frac{1}{{\tau  \uparrow }} = \frac{4}{3}\frac{{\nu {e^4}m}}{{\pi {\hbar ^3}{\kappa ^2}}}\frac{1}{{\left( {1 - s} \right)}}\int\limits_0^1 {dQ\frac{1}{{Q{{\left[ {1 + \left( {a/2} \right)\alpha \left( {Q,0} \right) + \left( {b/2} \right)\alpha \left( {\left( {b/a} \right)Q,0} \right)} \right]}^2}}}} ,\\
\frac{1}{{\tau  \downarrow }} = \frac{4}{3}\frac{{\nu {e^4}m}}{{\pi {\hbar ^3}{\kappa ^2}}}\frac{1}{{\left( {1 + s} \right)}}\int\limits_0^1 {dQ\frac{1}{{Q{{\left[ {1 + \left( {b/2} \right)\alpha \left( {Q,0} \right) + \left( {a/2} \right)\alpha \left( {\left( {a/b} \right)Q,0} \right)} \right]}^2}}}}. 
\end{array}
\label{equ18}
\end{equation}

We also need the modified $\beta $-functions. For spin up electrons we have
\begin{equation}
\begin{array}{l}
\beta  \uparrow  = \frac{{ya}}{{\nu \pi }}\left\{ {\left[ {1 - \int\limits_0^1 {dQ\frac{1}{{Q\left[ {1 + \left( {a/2} \right)\alpha \left( {Q,0} \right) + \left( {b/2} \right)\alpha \left( {\left( {b/a} \right)Q,0} \right)} \right]}}} } \right.} \right.\\
 + \frac{1}{\pi }\int\limits_0^\infty  {dQ\int\limits_0^\infty  {dW\left[ {\frac{1}{{1 + \frac{a}{2}\alpha \left( {Q,iWQ} \right) + \frac{b}{2}\alpha \left( {\frac{b}{a}Q,i{{\left( {\frac{b}{a}} \right)}^2}WQ} \right)}} - 1} \right]} } \\
\left. {\quad \quad  \times \left[ {\ln \left| {\frac{{{W^2} + {{\left( {Q + 1} \right)}^2}}}{{{W^2} + {{\left( {Q - 1} \right)}^2}}}} \right| - \frac{{2\left( {Q + 1} \right)}}{{{W^2} + {{\left( {Q + 1} \right)}^2}}} - \frac{{2\left( {Q - 1} \right)}}{{{W^2} + {{\left( {Q - 1} \right)}^2}}}} \right]} \right\}\\
 + \frac{{2{y^2}{a^5}}}{{3\nu {\pi ^2}}}\int\limits_0^\infty  {dQ\frac{1}{{{Q^2}\left( {{Q^2} - 1} \right){{\left[ {1 + \left( {a/2} \right)\alpha \left( {Q,0} \right) + \left( {b/2} \right)\alpha \left( {\left( {b/a} \right)Q,0} \right)} \right]}^2}}}} \\
\quad \quad \quad \quad \quad \quad  \times \left[ {1 - \frac{{{Q^2} - 1}}{{2Q}}\ln \left| {\frac{{Q + 1}}{{Q - 1}}} \right|} \right].
\end{array}
\label{equ19}
\end{equation}
To get $\beta  \downarrow $ one just interchanges $a$ and $b$ in Eq.\,(\ref{equ19}).

Now we have all formalism needed for the calculation of the magnetoresistance, $\Delta \rho /\rho  = \left[ {\rho \left( s \right) - \rho \left( 0 \right)} \right]/\rho \left( 0 \right)$, as a function of spin polarization $s$. The experiments are not given as a function of $s$ but as a function of {\bf B}. If the fields are small enough we can assume a linear relation between {\bf B} and $s$. It can be written as
\begin{equation}
B\left[ T \right] = \frac{{2.099879 \times {{10}^{ - 11}}{{\left( {n\left[ {c{m^{ - 3}}} \right]/\nu } \right)}^{2/3}}}}{{{m_{de}}\left( {\chi /{\chi _0}} \right)}}s.
\label{equ20}
\end{equation}

\section{\label{Res}Experimental and theoretical results}

Our theoretical and experimental results are compared in  Figs.\,\ref{figu2} - \ref{figu6}.  We have adjusted the spin-susceptibility enhancement-factor ${\left( {\chi /{\chi _0}} \right)}$ appearing in Eq.\,(\ref{equ20}) to get a reasonable fit between the theoretical and experimental curves. The adjustment only affects the theoretical curves in the horizontal direction. The enhancement factor is the only fitting parameter we have introduced. The enhancement of the density of states at the Fermi level increases when the density comes closer to ${n_c}$ ($n_c=3.5-4.4\times10^{18} \rm{cm}^{-3}$\,\cite{Ferr}) from the metallic side. It is furthermore well known that the spin susceptibility $\chi $ is more and more enhanced the closer to ${n_c}$ one gets and that the enhancement is reduced when the temperature goes up\,\cite{QM1,QM2,QM3,Ferr2}. Our extracted ${\left( {\chi /{\chi _0}} \right)}$ comply with this behavior. In the figures the upper horizontal axes show the spin polarization parameter, $s$, defined in Eq.\,(\ref{equ13}). The relation between $s$ (upper horizontal axis) and $B$ (lower horizontal axis) is given by Eq.\,(\ref{equ20}). 

\begin{figure}
\includegraphics[width=8.0cm]{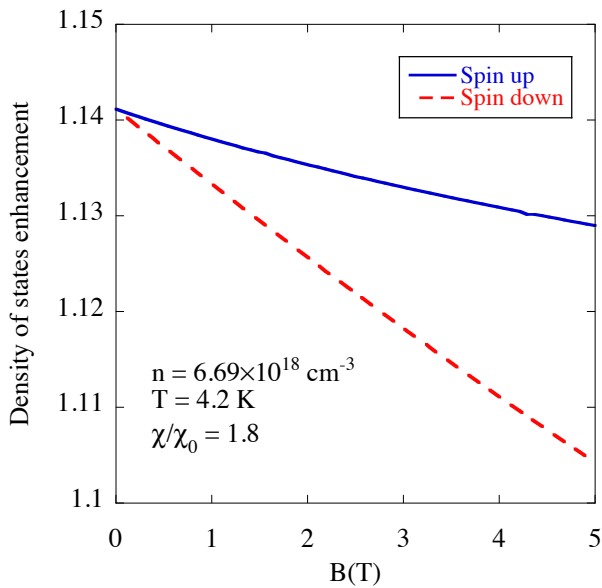}
\caption{(Color online) The enhancement of the density of states at the Fermi-level for both spin types as function of magnetic induction, B. The results are vaiid for the doping concentration $6.69\times10^{18}$ cm$^{-3}$ at $4.2$ K. }
\label{figu7}
\end{figure}

Now, what causes the negative magnetoresistance? As we mentioned above the density of states is enhanced at the Fermi-level in absence of a magnetic field. This leads to an enhancement of the resistivity.  In absence of magnetic field the spin up and spin down bands are degenerate and the Fermi wave-numbers are the same for both spin types. When the magnetic field is introduced the spin down bands move down in energy and the spin up bands move up. There is a redistribution of electrons from the spin up bands to the spin down bands so that the Fermi-level is the same in all bands. The Fermi wave-numbers are now different in the two band types. See the inset of Fig.\,\ref{figu2}. When the magnetic field is introduced the density of states of both electron types, i.e. spin up and spin down electrons, are enhanced for states with wave-number ${k_0} \uparrow $ and ${k_0} \downarrow $. This means that the peak in the density of states at the Fermi-level is for each spin type split up into two. The states involved in the enhancement of the density of states are indicated by circles in the inset of Fig.\,\ref{figu2}. For spin up electrons one peak remains at the Fermi-level while the other moves up into the unoccupied part of the bands. For spin down electrons one peak remains at the Fermi-level and one moves further down in the occupied part of the bands. For both spin types the enhancement at the Fermi-level is hence reduced. It is only the enhancement at the Fermi level that effects the resistivity. This causes the initial negative magnetoresistance. The enhancement of the density of states at the Fermi-level for both spin types as function of magnetic induction is shown in Fig.\,\ref{figu7} for the sample with doping concentration $6.69\times10^{18}$ cm$^{-3}$ at $4.2$ K.

There is another effect that acts in the same direction. There are Friedel oscillations\,\cite{Friedel} in the screening-charge density centered around each impurity potential with Fourier component $q = 2{k_0}$ or $Q = 1$. This leads to an enhanced scattering rate in the back scattering direction across the Fermi spheres and an enhancement of the resistivity. At zero magnetic field the Friedel oscillations have the same periodicity for spin up and spin down electrons; both type of electrons scatter equally strongly against both Friedel oscillations. When the magnetic field is turned on the Friedel oscillations will split up into two; one with Fourier component  $q=2{k_0}\uparrow $; one with Fourier component $q=2{k_0}\downarrow $. This means that  the back-scattering rate for an electron of a certain spin against the Friedel oscillations of the opposite spin is reduced. The enhancement of the resistivity is thus reduced leading to a negative magnetoresistance.  However, this effect is expected to give a much smaller contribution to the negative magnetoresistance than the density of states effect since the electron can scatter with a wave number ranging from zero up to two times the Fermi wave number.

All our calculations are for zero temperature. What happens for non-zero temperatures? If we study classical experiments\,\cite{Yam} we find that the negative magnetoresistance effect is  gradually reduced when the temperature is enhanced. This is consistent with our theory. The peak at the Fermi-level of the density of states is expected to be broadened. Besides, at zero temperature only states at the Fermi-level takes part in the conductivity. When the temperature goes up also states away from the Fermi-level where the enhancement in the density of states is weaker take part. Both these effects are expected to gradually remove the negative magnetoresistance. The temperature effects are expected to be more and more important the lower the density. This is also what is found experimentally\,\cite{Yam}.

\begin{figure}
\includegraphics[width=8.0cm]{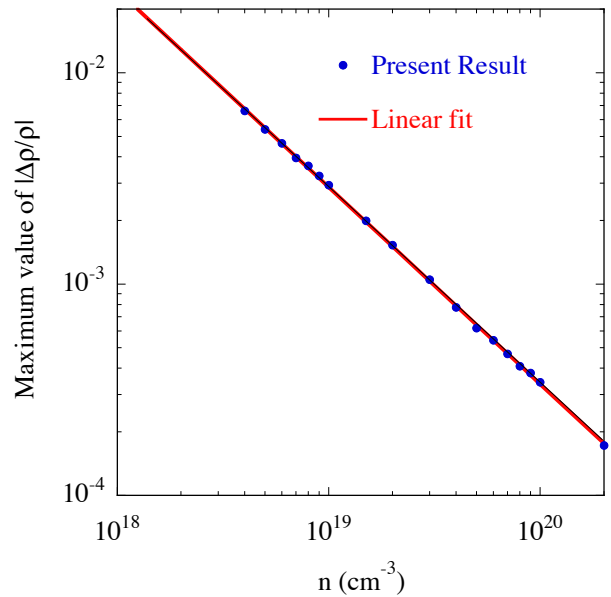}
\caption{(Color online) The modulus of the maximum negative magnetoresistance from our theoretical calculations. It shows linear relation on a log-log plot which means a power law dependence on the doping concentration .}
\label{figu8}
\end{figure}

In Fig.\,\ref{figu8} we see that the maximum negative magnetoresistance increases linearly on a log-log plot when the density is reduced. For finite temperature the maximum is expected to start decreasing at a density that depends on the temperature. The higher the temperature the earlier the decrease is expected to set in. In Fig.\,8 of Ref.\,\cite{Yam} this decrease is observed.

\section{\label{Sum}Summary and conclusions}

We have performed magnetoresistance measurements of heavily phosphorous doped silicon and compared the results to theory. The resistance was calculated using the so-called generalized Drude approach taking many-body effects into account. We propose that the origin of the 
negative magnetoresistance is a combination of two effects. The many-body effects lead to an enhancement of the density of states at the Fermi level which in turn results in an enhancement of the resistivity. Friedel oscillations in the screening charge density cause an enhanced back-scattering rate across the Fermi volumes leading to an additional enhancement of the resistivity.

When the magnetic field is turned on the 
enhancement of the density of states for each spin type is split up into two peaks, one at the Fermi-level and one that moves away from the Fermi-level with enhanced magnetic field. This reduces the resistivity. Also, the back-scattering rate against the Friedel oscillations will be reduced in the presence of a magnetic field. Both these effects act towards a negative magnetoresistance. The first effect is expected to be dominating. 
We would have liked to have samples that cover a broader range of doping concentrations from closer to $n_c$ to a much higher value. Unfortunately, we have had contact problems preventing high quality measurements. We hope to come back with complementary results in the near future. 

From our work we can deduce that the metal-non-metal transition in heavily $n$-doped Si is driven by electron-electron and electron doping-ion interactions and not by spin-flip scattering or spin-orbit scattering. This is in line with the findings in Ref.\,\cite{Sara} for $p$-doped Si (Si:B) where these conclusions could be drawn from the universal scaling of the magnetoconductance of that system.

We have not been able to find any recent work on the magnetoresistance of heavily doped Si near the metal-non-metal transition. We hope that the present work will inspire the readers to perform new such measurements. Recently\,\cite{Wink} one has managed to produce samples of sulfur-doped silicon at non-equilibrium concentrations in a range covering the concentration of the metal-non-metal transition. This was achieved by using ion implantation followed by pulsed-laser melting and rapid resolidification. Sulfur is a deep level impurity in crystalline silicon and $n_c$ is much higher than for phosphorus doping. One estimated $n_c$ to be between 1.8 and $4.3 \times {10^{20} }{\rm{c}}{{\rm{m}}^{ - 3}}$, i.e., two orders of magnitude higher than for phosphorous doping. It would be very interesting to see magnetoresistance measurements being performed on these samples.

\acknowledgments
AFdS, AL, ZSM, and HB acknowledge financial support of the Brazilian agencies CNPq, FAPESB/PRONEX,  and FAPESP.


\begin{thebibliography}{10}

\bibitem{Kel}W. Thomson, Proc. Royal Soc. London {\bf 8} 546 (1857)
\bibitem{Fert}  M. N. Baibich, J. M. Broto,  A. Fert,  F. Nguyen Van Dau, F. Petroff,  P.  Etienne, G. Creuzet,  A. Friederich,  and J. Chazelas, Phys. Rev. Lett. {\bf  61}, 2472 (1988)
\bibitem{Grun}G. Binasch, P. Gr\"unberg, F. Saurenbach, and W. Zinn,  Phys. Rev. B {\bf 39} 4828 (1989). 
\bibitem{Jonk}G. H. Jonker, and J. H. Van Santen,  Physica {\bf 16}, 337(1950).
\bibitem{Ram}  A. P. Ramirez,  Journal of Physics: Condensed Matter {\bf 9}, 8171(1997).
\bibitem{Jull}M. Julliere, Phys. Lett. {\bf 54A}, 225 (1975).
\bibitem{Sol}S. A Solin, T. Thio, D. R. Hines,  J. J. Heremans, Science {\bf 289}, 1530 (2000),
\bibitem{Ali} M. N. Ali, J. Xiong,	S. Flynn, J. Tao, Q. D. Gibson,	L. M. Schoop,	T. Liang,	N. Haldolaarachchige, M. Hirschberger,	N. P. Ong,	
and R. J. Cava, Nature {\bf 514}, 205 (2014).

\bibitem{Yam} C. Yamanouchi, K. Miziguchi, and W. Sasaki, J. Phys. Soc. Jpn. {\bf 22}, 859 (1967), and references therein.
\bibitem{Sas}  W. Sasaki, J. Phys. Soc. Jpn. {\bf 21}, 543 (1966).
\bibitem{Ion} A. N. Ionov, I. S. Shlimak, and A. L. Efros, Sov. Phys. Solid State {\bf 17}, 1835 (1976).
\bibitem{And} D. G. Andrianov, G. U. Lazareva, A. S. Savel'ev, and V. I. Fistul', Sov. Phys. Semicond. {\bf 9}, 141 (1975).
\bibitem{Zav} E. I. Zavitskaya, I. D. Voronova, and N. V. Roshdestvenskaya, Sov. Phys. Semicond. {\bf 6}, 1668 (1973).
\bibitem{Emel} 0. V. Emel'yanenko, T. S. Lagunova, K. G. Masagutov, D. N. Nasledov, and D. D. Nedeoglo, Sov. Phys. Semicond. {\bf 9}, 1001 (1976).
\bibitem{Ish} S. Ishida and E. Otsuka, J. Phys. Soc. Jpn. {\bf 42}, 542
(1977).
\bibitem{Kho} B. P. Khosla and J. R. Fisher, Phys. Bev. B {\bf 2}, 4084 (1970).
\bibitem{Toy}Y. Toyozawa, J. Phys. Soc. Jpn. {\bf 17}, 986 (1962)

\bibitem{Alex}M.  N. Alexander, and D. F. Holcomb, Rev. Mod. Phys. {\bf 40}, 815 (1968).

\bibitem{SerBer}B. E. Sernelius, and K.-F. Berggren, Phil. Mag. {\bf 43} 115 (1981).

\bibitem{John}H. G. Johnson, S. P. Bennett, R. Barua, L. H. Lewis, and D. Heiman, Phys. Rev. B {\bf 82}, 085202 (2010).
\bibitem{Por}N. A. Porter, and  C. H. Marrows, Sci Rep. 2012;2:565

\bibitem{Bran} W. R. Branford, A. Husmann, S. A. Solin, S. K. Clowes, T. Zhang, Y. V. Bugoslavsky, and L. F.
Cohen, Appl. Phys. Lett. {\bf 86}, 202116-1 (2005).
\bibitem{Par1} M. M. Parish, and P. B. Littlewood, Nature (London) {\bf 426}, 162 (2003).
\bibitem{Par2} M. M. Parish,  and P. B. Littlewood, Phys. Rev. B {\bf 72}, 094417 (2005).
\bibitem{Hu} J. Hu, M. M. Parish, and T. F Rosenbaum, Phys. Rev. B {\bf 75}, 214203 (2007).
\bibitem{Abr1} A.A. Abrikosov, Phys. Rev. B {\bf 58}, 2788 (1998).
\bibitem{Abr2}A.A. Abrikosov, Europhys. Lett. {\bf 49}, 789 (2000).
\bibitem{Gut1} V. Guttal and D. Stroud, Phys. Rev. B {\bf 71}, 201304(R) (2005).
\bibitem{Gut2}V. Guttal and D. Stroud, Phys. Rev. B {\bf 73}, 085202 (2006).
 

%
\bibitem{Zieg} J.  F. Ziegler, J. P. Biersak, and U. Littmark, {\it The Stopping and Ranges of Ion in Solids} Vol. I (Pergamon, New York, 1985).

\bibitem{Pauw} L. J. Van der Pauw, {\it Philips Res. Rep.} {\bf 13}, 1 (1958).
\bibitem{Ferr} A. Ferreira da Silva, Bo E. Sernelius, J. P. de Souza, H. Boudinov, H. Zheng, and M. P. Sarachik, Phys. Rev. B {\bf 60}, 15824 (1999).
\bibitem{Abram} E. Abramof, A. Ferreira da Silva, Bo E. Sernelius, J. P. de Souza, and H. Boudinov,  Phys. Rev. B {\bf 55}, 9584 (1997).


\bibitem{Ser1} Bo E. Sernelius, Phys. Rev. B {\bf 41}, 3060 (1990).
\bibitem{Ser2} Bo E. Sernelius, and K.-F. Berggren, Phys. Rev. B {\bf 19}, 6390 (1979).
\bibitem{SerMor}Bo E. Sernelius, and M. Morling, in {\it Shallow Impurities in Semiconductors 1988}, Inst. Phys. Conf. Ser. No 95, edited by B. Monemar (IOP, Bristol, 1989), p. 555.
\bibitem{Ser3} Bo E. Sernelius, Phys. Rev. B {\bf 40}, 12438 (1989).
\bibitem{Ser4} Bo E. Sernelius, Phys. Rev. B {\bf 43}, 7136 (1991).
\bibitem{Ziman} J. M. Ziman, Phil. Mag. {\bf 6}, 1013 (1961).
\bibitem{QM1} J. D. Quirt and J. R. Marco, Phys Rev. Lett. {\bf 26}, 318 (1971).
\bibitem{QM2} J. D. Quirt and J. R. Marco, Phys. Rev. B {\bf 7}, 3842 (1973).
\bibitem{QM3} J. D. Quirt and J. R. Marco, Phys. Rev. B {\bf 5}, 1716 (1972).
\bibitem{Ferr2} A. Ferreira da Silva, Phys. Rev. B {\bf 38}, 10 055 (1988).
\bibitem{Friedel} J. Friedel, Nuovo Cimento Suppl. {\bf 7}, 287 (1958).


\bibitem{Sara}S. Bogdanovich, P. Dai, M. P. Sarachik, and V. Dobrosavljevic, Phys. Rev. Lett.{\bf 74}, 2543 (1995).
\bibitem{Wink} M. T. Winkler, D. Recht, M.-J. Sher, A. J. Said, E. Mazur, and M. J. Aziz, Phys. Rev. Lett. {\bf 106}, 178701 (2011).


\end{thebibliography}
\end{document}